\documentclass[a4paper,11pt]{article}
\usepackage{pos}

\title{MSHT Approximate N3LO PDFs: Updates and Consequences for Phenomenology}

\author*[a]{T. Cridge}
\author[b]{L. A. Harland-Lang}
\author[b]{R. S. Thorne}

\affiliation[a]{Elementary Particle Physics, University of Antwerp,\\
  Groenenborgerlaan 171, 2020 Antwerp, Belgium}

\affiliation[b]{Department of Physics and Astronomy, University College London,\\
London, WC1E 6BT, UK}

\emailAdd{thomas.cridge@uantwerpen.be}

\abstract{
We present updates to the MSHT approximate N3LO PDFs focusing upon recent developments, examining the impacts of newly determined splitting function and transition matrix element calculations, performed since the public MSHT20aN3LO PDF set was released. We observe only small changes to the output PDFs, at most of similar size to their quoted uncertainties and typically notably less. Finally, we also present the impacts on the PDF luminosities, where changes are generally further reduced.}

\FullConference{XXXII International Workshop on Deep Inelastic Scattering and Related Subjects (DIS2025)\\
24-28 March, 2025\\
Cape Town, South Africa\\}

\begin{document}
\maketitle

\section{Introduction}
\label{sec:intro}

The improvement in accuracy and precision in theoretical calculations required by the LHC in recent years 
 motivated our work in the MSHT PDF group to produce the world's first approximate N3LO (aN3LO) PDFs \cite{McGowan:2022nag} including also estimates for the missing higher order uncertainties (MHOUs): {\tt MSHT20aN3LO}. In this work we included the theoretical ingredients for the splitting functions~\cite{Catani:1994sq, Vogt:2018miu,Moch:2021qrk,vanNeerven:1999ca,vanNeerven:2000uj,Moch:2017uml,Moch:2004pa,Lipatov:1976zz,Kuraev:1977fs,Balitsky:1978ic,Fadin:1998py,Ciafaloni:1998gs}, transition matrix elements~\cite{Buza:1996wv,Buza:1996xr,Kawamura:2012cr,Bierenbaum:2009mv,Ablinger:2014vwa,Blumlein:2021enk,Ablinger:2014nga,ablinger:agq,Blumlein:2017wxd,Ablinger:2014uka,Ablinger:2014tla}, and structure function coefficient functions~\cite{Catani:FFN3LO1,Laenen:FFN3LO2,Kawamura:2012cr,Vermaseren:2005qc} available at the time. However, in the period since there has been further work, increasing our knowledge of the former two ingredients, with an increase in the number of Mellin moments known for several of the splitting functions~\cite{Falcioni:2023luc,Falcioni:2023vqq,Falcioni:2023tzp,Falcioni:2024xyt,Falcioni:2024qpd}, and calculations of transition matrix elements~\cite{Ablinger:2022wbb,Ablinger:2023ahe,Ablinger:2024xtt} that previously could only be approximated. In these proceedings we summarise the impact of updating the MSHT20aN3LO PDFs incorporating these additional theory inputs. This builds on top of our various other subsequent studies at approximate N3LO~\cite{Jing:2023isu,Cridge:2023ryv,Cridge:2023ozx,Cridge:2024exf,Thorne:2024npj,Thorne:2024ywy}. In Section~\ref{sec:Theory} we outline the new information included, and in Section~\ref{sec:PDFs} we describe the impacts on the PDFs and on phenomenology via the PDF luminosities. We then conclude in Section~\ref{sec:conclusions}.

\section{Theory Updates}
\label{sec:Theory}

In this work we include updates on the splitting functions and transition matrix elements used in the MSHT approximate N3LO PDFs. The resulting changes are:
\begin{itemize}
\item Splitting Functions - For the singlet splitting functions $P_{ij}$, where $ij= qq^{PS},qg,gq,gg$ there are now 10 Mellin moments rather than 4 available previously~\cite{Falcioni:2023luc,Falcioni:2023vqq,Falcioni:2023tzp,Falcioni:2024xyt,Falcioni:2024qpd}. (For the non-singlet $P_{qq}^{NS}$ there is no change - we have 8 Mellin moments as in MSHT20aN3LO based on \cite{Moch:2017uml}). We use the central values of the updated splitting function approximations provided directly by the authors (``FHMRUVV'') for these updates.
\item Transition Matrix Elements (``TMEs'') - For the transition matrix elements we include new calculations of $A_{gg,H}$ and $A_{Hg}$~\cite{Ablinger:2022wbb,Ablinger:2023ahe,Ablinger:2024xtt}, which were previously approximated. In addition, $A_{qq,H}$ was previously approximated in our work and is now included exactly based on~\cite{Ablinger:2014vwa,Ablinger:2010ty}, whilst we also add $A_{qg,H}$ \cite{Ablinger:2010ty} for completeness, though it is a very small contribution. We include these updates via grids of the matrix elements provided by our NNPDF colleagues in the context of the approximate N3LO combination work~\cite{MSHT:2024tdn}.
\end{itemize}

We note that the benchmarking work carried out in \cite{Cooper-Sarkar:2024crx} involved as part of the study updating the PDF evolution to the splitting functions available at the time~\cite{Falcioni:2023luc,Falcioni:2023vqq,Falcioni:2023tzp,Moch:2023tdj}, in that case $P_{qq}^{PS}$ and $P_{qg}$ were updated to the 10 Mellin moments used here but $P_{gq}$ and $P_{gg}$ only had 5 Mellin moments\footnote{$P_{gq}$ and $P_{gg}$ were updated in that work by an additional Mellin moment (to 5) relative to the original MSHT20aN3LO study (which used 4) due to private communications from the authors.}. The impacts of these changes in the splitting functions were reported in \cite{Thorne:2024npj,Thorne:2024ywy}. These are supplemented with the further updates to $P_{gq}$ and $P_{gg}$ here~\cite{Falcioni:2024xyt,Falcioni:2024qpd}, which now have 10 Mellin moments calculated and utilised.

\begin{figure}
\begin{center}
\includegraphics[scale=0.23]{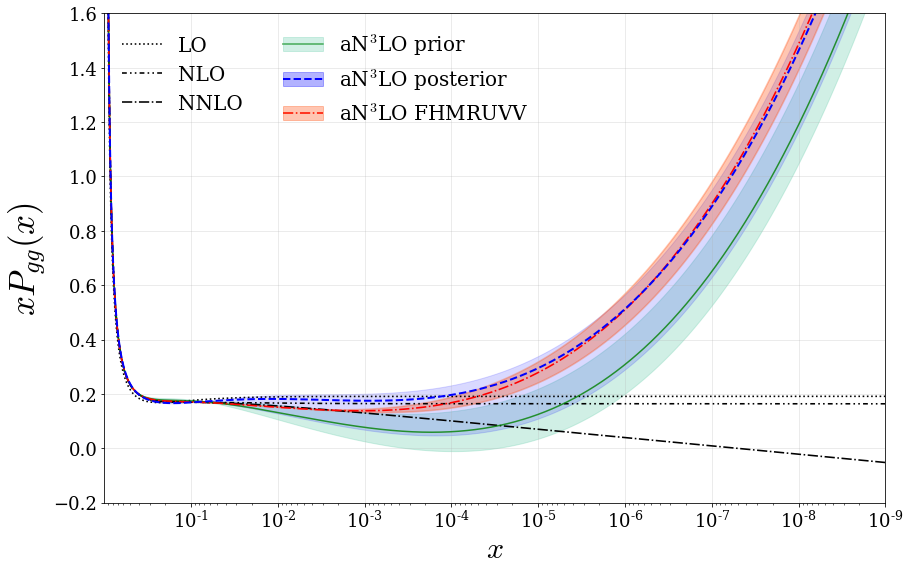}
\includegraphics[scale=0.23]{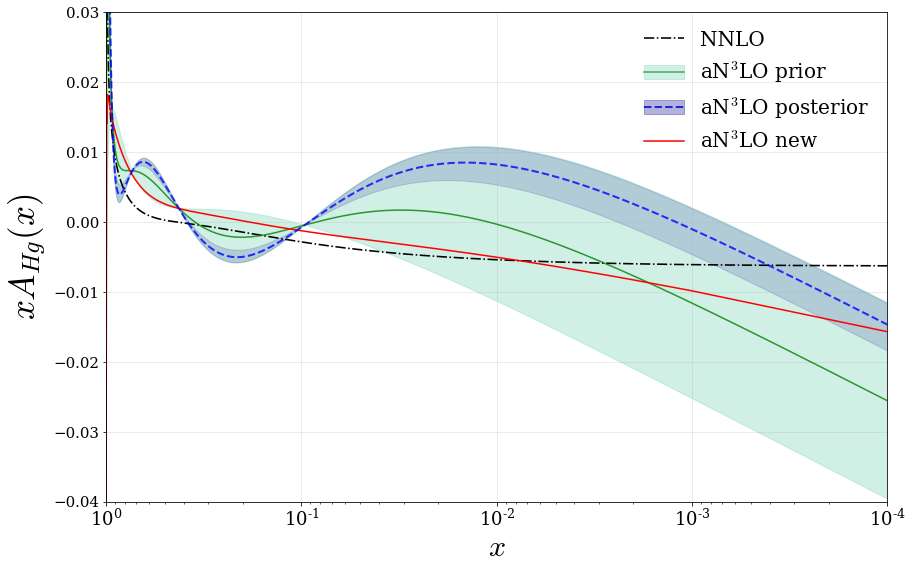}
\end{center}
\caption{Comparison of our previous approximations of (a) $P_{gg}$ and (b) $A_{Hg}$ - with our prior/posterior determination in MSHT20aN3LO~\cite{McGowan:2022nag} shown in green/blue with associated error bands. In (a) the new updated approximate determination of FHMRUVV~\cite{Falcioni:2024qpd} is shown and in (b) the new updated parameterised version of the exact determination of~\cite{Ablinger:2023ahe,Ablinger:2024xtt} is shown, both in red.}
\label{fig:PggandAHg_updated}
\end{figure}

Before showing the resulting PDFs we first show in Figure~\ref{fig:PggandAHg_updated} example comparisons, showing in Figure~\ref{fig:PggandAHg_updated}(a) the updated $P_{gg}$ splitting function relative to our previous approximations, and in Figure~\ref{fig:PggandAHg_updated}(b) a similar comparison for the updated $A_{Hg}$. Beginning with $P_{gg}$, we observe that the agreement between the new result for $P_{gg}$~\cite{Falcioni:2024qpd} of FHMRUVV with our aN3LO posterior (which is that ultimately used for the output PDFs) is very good\footnote{We note that in the MSHT20aN3LO paper~\cite{McGowan:2022nag} the $\rho_{gg}$ variation was stated as $-5 < \rho_{gg} < 15$, this was a typo and should read $5 < \rho_{gg} < 15$.}, within its uncertainty band over the entire $x$ range. In particular, we note that the agreement is closer to our posterior than prior approximation. As for the transition matrix element $A_{Hg}$, whilst the agreement is not quite as good as for the splitting function $P_{gg}$, the new result of \cite{Ablinger:2023ahe,Ablinger:2024xtt} is within the uncertainty band of our prior approximation. Whilst our prior, and more so our posterior approximation, are more oscillatory than the new result, such oscillation will be diluted and wash out in the convolution to form the PDFs.

\section{Impact on the PDFs}
\label{sec:PDFs}

\begin{figure}[htbp!]
\begin{center}
\includegraphics[height=4.25cm,width=6cm,trim=0.3cm 0cm -0.0cm 0cm,clip]{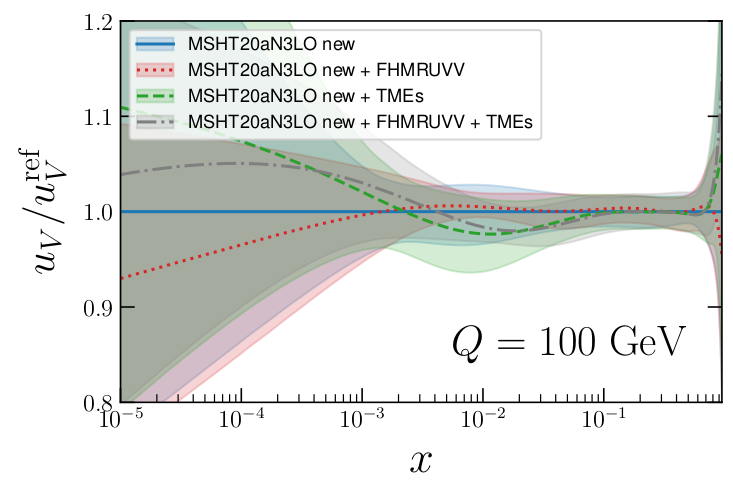}
\includegraphics[height=4.25cm,width=6cm,trim=0.2cm 0cm -0.2cm 0cm,clip]{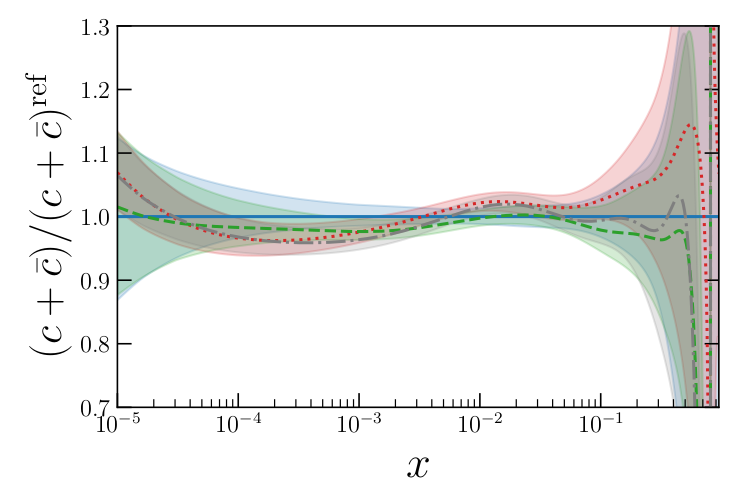}
\hfill \\
\includegraphics[height=4.25cm,width=6cm]{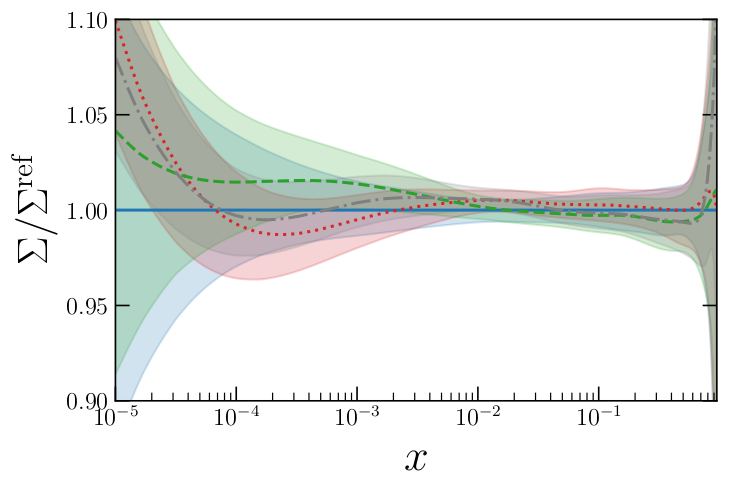}
\includegraphics[height=4.25cm,width=6cm]{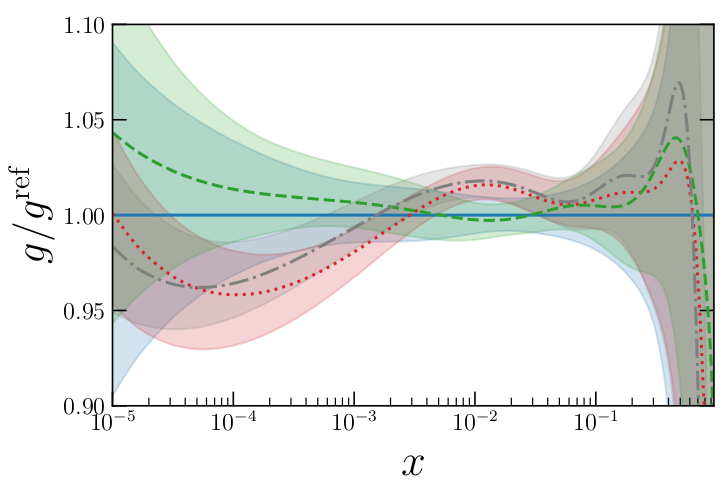}
\end{center}
\vspace{-0.35cm}
\caption{Comparison of the updated PDFs with only the new FHMRUVV splitting functions (red dotted), only the new transition matrix elements (green dashed), and both of these updates (grey dash-dotted), against those with our original approximations for these pieces (blue)~\cite{McGowan:2022nag}. Note the latter is described as ``MSHT20aN3LO new'' as it has some updates in the data and methodology relative to the public set, as described in \cite{Cridge:2023ryv,Cridge:2023ozx}.}
\label{fig:PDFs_updated}
\end{figure}

The main purpose of this work is to investigate the impact of these theoretical updates on the output PDFs. This is shown in Figure~\ref{fig:PDFs_updated}, where we show the up valence, total charm, total singlet and gluon PDFs and we provide the impacts of updating the splitting functions or transition matrix elements separately as well as their combination. It should be noted that whilst the uncertainties on the PDFs are shown, as we utilise the results of ~\cite{Falcioni:2023luc,Falcioni:2023vqq,Falcioni:2023tzp,Falcioni:2024xyt,Falcioni:2024qpd,Ablinger:2022wbb,Ablinger:2023ahe,Ablinger:2024xtt,Ablinger:2014nga,Bierenbaum:2009mv,Ablinger:2014vwa,Blumlein:2021enk,Ablinger:2010ty} directly, we have not yet incorporated an estimate of the MHOU uncertainty remaining at N3LO and beyond in the splitting functions, or at N$^{4}$LO for the transition matrix elements, into our fits. These would be required to completely capture the MHOU theoretical uncertainty in our nuisance parameter approach. Whilst there would be expected impacts on the fit qualities from incorporating these additional uncertainties into the fit, the effect on the PDFs is anticipated to be small and will be checked in a future more complete update.

Figure~\ref{fig:PDFs_updated} shows that all PDFs have only minor changes upon these theory updates, largely within the uncertainty bands of the original PDFs. The up valence, charm and total singlet PDFs remain within the uncertainty band of the MSHT20aN3LO PDFs across the whole $x$-range. There is a small increase of $\approx 2\%$ in the charm PDF at $x \approx 10^{-2}$ from the use of the FHMRUVV splitting functions (red dotted) relative to the original case which mimics a similar small effect seen in the PDF evolution benchmarking study~\cite{Cooper-Sarkar:2024crx}\footnote{In the PDF evolution benchmarking the charm PDF had a similar magnitude dip at this location from the PDF evolution alone relative to our previous approximation, we can expect that in a fit in the data region this may be compensated for in the PDFs resulting in a similar magnitude increase in the charm PDF as observed here.}. The charm PDF does have larger changes near threshold (not shown), where the transition matrix elements have a larger impact and are not washed out by the evolution, though at $Q=1.9~{\rm GeV}$ the PDFs are still consistent above $x \approx 10^{-2}$. The total singlet, $\Sigma$, shows differences of at most $1\%$ down to $x \sim 10^{-4}$, consistent with the evolution benchmarking study~\cite{Cooper-Sarkar:2024crx}, at which point the the FHMRUVV splitting functions cause a rise at low $x$, albeit within the growing uncertainty band down to $x \sim 10^{-5}$.

Finally, in the bottom right of Figure~\ref{fig:PDFs_updated} we show the gluon, which is of particular phenomenological importance given its connections with Higgs production. Our approximate N3LO PDFs showed a dip of up to approximately 5\% in the gluon at $x \approx 10^{-2}$ relative to NNLO, the more recent NNPDF approximate N3LO PDFs also show a similar feature, albeit reduced in magnitude~\cite{NNPDF:2024nan}. Upon updating our PDFs with additional moments for $P_{qq}^{PS}$ and $P_{qg}$ (the full 10 Mellin moments) and $P_{gq}$ and $P_{gg}$ (only the 5 Mellin moments known at the time) we reported in \cite{Thorne:2024npj,Thorne:2024ywy} that the gluon PDF dip was slightly reduced, rising by about 1.5\% near $x \approx 10^{-2}$ (again reflecting what was seen in the evolution benchmarking study~\cite{Cooper-Sarkar:2024crx}). Upon the further update of the Mellin moments of $P_{gq}$ and $P_{gg}$ to the results of~\cite{Falcioni:2024xyt,Falcioni:2024qpd}, there are only small further differences with a total rise of less than 2\% near $x \approx 10^{-2}$ observed relative to the baseline. The gluon is then reduced below $x \approx 3 \times 10^{-3}$, similar to results seen in \cite{Hekhorn:2025xke}. Upon further updating of the transition matrix elements we observe a very slight further increase below $x \approx 10^{-2}$. The transition matrix elements additionally raise the gluon slightly, though within uncertainties, at large $x$. 

There are also impacts on the fit quality. We observe a worsening of around 40 points in $\chi^2$ upon incorporation of the updated FHMRUVV splitting functions, relative to identical fits with our original approximations for the splitting functions\footnote{The settings of the PDFs used here are updated relative to the public aN3LO PDFs by changes noted in \cite{Cridge:2023ozx,Cridge:2023ryv} though without the ATLAS 8~TeV jets here for a more direct comparison.}. This is still an improvement of almost 180 points in $\chi^2$ relative to NNLO. Adding on top of this the updated transition matrix elements there is a further worsening of the fit quality by almost 90 points, leaving the aN3LO fit quality still around 90 units better than NNLO. The future incorporation of the remaining MHOU uncertainties would also allow the fit to improve somewhat further.  The differences in fit quality for the splitting function updates are focused at small $x$, similar to those reported in \cite{Thorne:2024npj,Thorne:2024ywy}, with the HERA data showing a worsening of approximately 50 units in $\chi^2$. Upon the further addition of the transition matrix element updates the changes are more distributed amongst datasets. The charm and bottom structure function data from HERA worsen by approximately 30 points in $\chi^2$ and the remainder of the HERA data by approximately 20 units, this leaves the total HERA data fit by approximately 20 units in $\chi^2$ worse than at NNLO (the charm and bottom structure function data being around 45 units worse fit than at NNLO whilst the rest of the HERA data is almost 25 units better than at NNLO). The fixed target data at high $x$ worsen by a total of 5 units, remaining significantly better fit than at NNLO by more than 15 units in $\chi^2$. Nonetheless, overall there is still a substantial preference for the N$^{3}$LO theory shown by the global fit data, including by the LHC Drell-Yan data in general and also by the ATLAS 8~TeV $Z$ $p_T$ data. As a result, the improvement in $\chi^2$ in going from NNLO to N3LO remains much larger than the number of additional theoretical nuisance parameters included (which also reduces as the ingredients become fully known and are updated).

\begin{figure}[htbp!]
\begin{center}
\includegraphics[height=4.25cm,width=6cm]{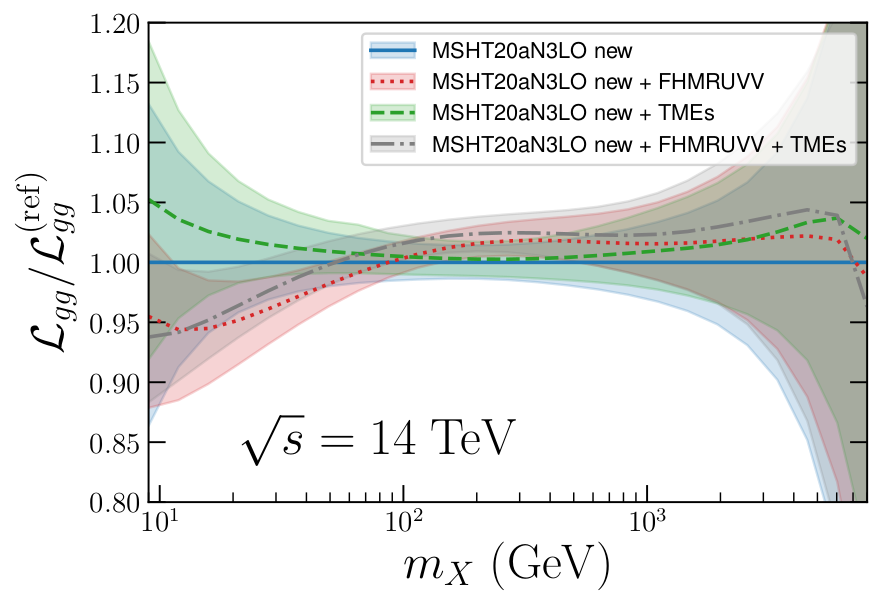}
\includegraphics[height=4.25cm,width=6cm]{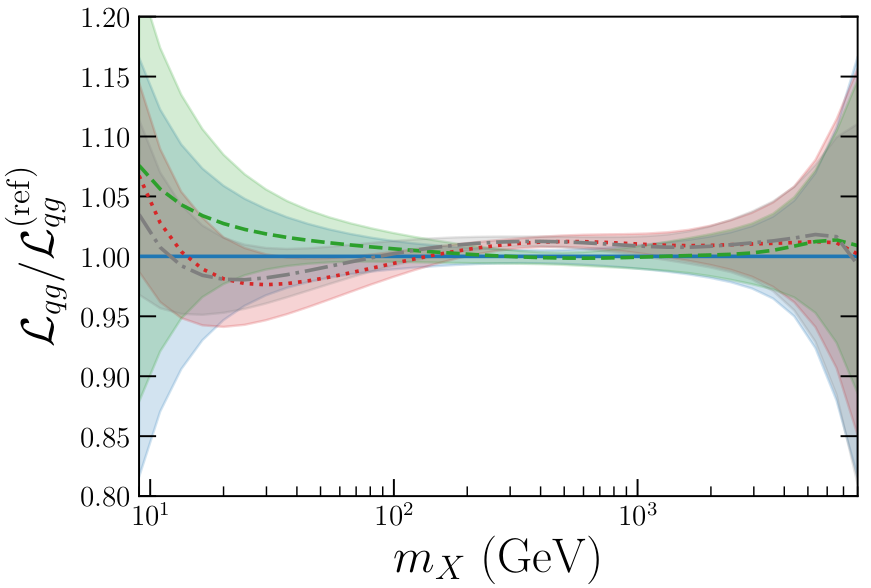}
\includegraphics[height=4.25cm,width=6cm,trim=-0.5cm 0cm 0.2cm 0cm,clip]{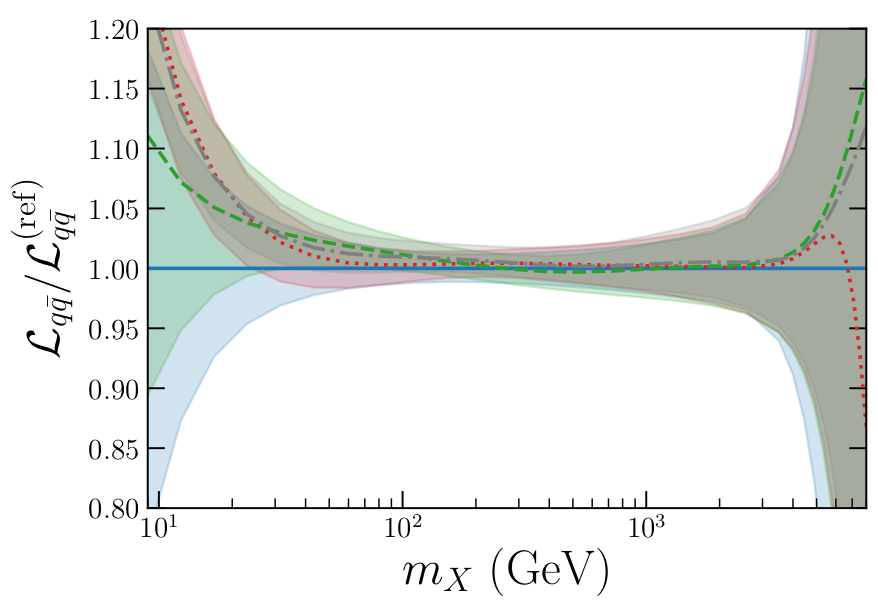}
\includegraphics[height=4.25cm,width=6cm]{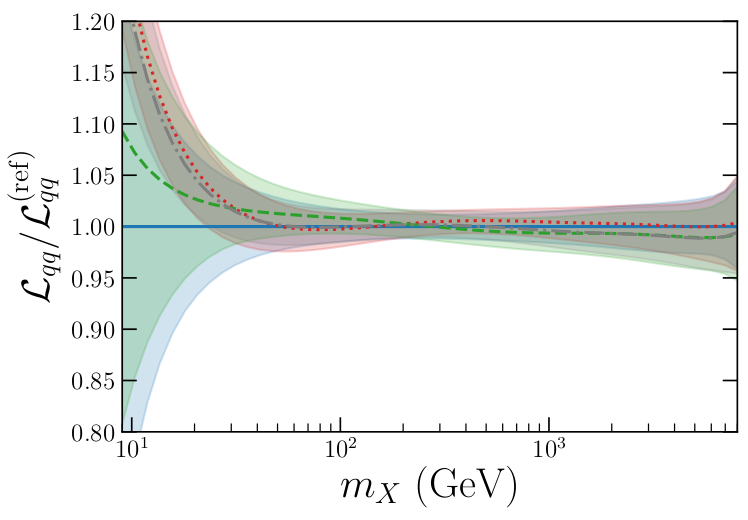}
\hfill%
\end{center}
\vspace{-0.35cm}
\caption{Comparison of the updated PDF luminosities at 14 TeV with only the new FHMRUVV splitting functions (red dotted), only the new transition matrix elements (green dashed), and both of these updates (grey dash-dotted), with our original approximations for these pieces (blue)~\cite{McGowan:2022nag}. Note the latter is described as ``MSHT20aN3LO new'' as it has some updates in the data and methodology relative to the public set, as described in \cite{Cridge:2023ryv,Cridge:2023ozx}.}
\label{fig:lumis_updated}
\end{figure}

Finally, we analyse the impact of these updates on the PDF luminosities, which are most directly connected to phenomenology. In Figure~\ref{fig:lumis_updated} we provide the gluon-gluon, quark-gluon, quark-antiquark and quark-quark luminosities. Beginning with the $gg$ luminosity in the top left we observe that (in the absence of rapidity cuts) the impact around $m_X = 125 {\rm GeV}$ is small, less than 1\% upon addition of the FHMRUVV splitting functions alone and less than 0.5\% upon instead adding the updated the transition matrix elements, the reduction in the $gg$ luminosity around the Higgs mass from NNLO to aN3LO with both sets of updates is therefore $\approx 3.5\%$. In particular, the impact of the updated FHMRUVV splitting functions is consistent with the update reported in \cite{Thorne:2024npj,Thorne:2024ywy}, where it was noted that the integration over $x$ washes out the small increase in the gluon PDF somewhat at the level of the $gg$ luminosity due to the reduction in the gluon below $x \approx 3 \times 10^{-3}$. Indeed, the impact of rapidity cuts was also analysed previously and found to change this little unless very stringent cuts were applied ($|y| < 0.4$). These cuts reduce the range of $x$ sampled in the PDFs and indicate that the most significant effects of the updates to the splitting functions are likely to be present in distributions rather than in the inclusive cross section. On the other hand, as noted in Section~\ref{sec:PDFs}, adding the transition matrix element updates on top of the splitting function updates raises the gluon slightly over a wide $x$ range below $x \approx 10^{-2}$, resulting in the observed further increase of 1\% in the $gg$ luminosity. The net effect is therefore a rise of around 2\% (or 1$\sigma$) in the $gg$ luminosity.

In fact, upon the addition of QED to aN3LO PDFs we demonstrated previously~\cite{Cridge:2023ryv} that the gluon reduces further relatively uniformly across $x$ due to the need to provide momentum to the photon PDF~\cite{Cooper-Sarkar:2025sqw}. As a result, a reduction of $\approx 6\%$ was observed in $\mathcal{L}_{gg}$ for the aN3LO+QED PDFs of \cite{Cridge:2023ryv} relative to NNLO. This QED effect on the gluon was observed to be approximately independent of the QCD order to which it is applied, whether QED is added to NNLO~\cite{Cridge:2021pxm} or aN3LO~\cite{Cridge:2023ryv}. Therefore whilst we have not performed this study, we anticipate adding QED on top of the updated aN3LO PDFs presented here, would further reduce the $gg$ luminosity by around $1\%$, though we leave the precise evaluation of this to future work. Such changes would then also impact the Higgs production cross-section in gluon fusion~\cite{MSHT:2024tdn,Cooper-Sarkar:2025sqw}. 

Of the other PDF luminosities shown in Figure~\ref{fig:lumis_updated}, the quark-quark and quark-antiquark luminosities are relatively unaffected with percent level differences across most of the invariant mass range and rising only at small invariant masses due to the rise in the quarks (as seen in the total singlet in Figure~\ref{fig:PDFs_updated}). The quark-antiquark luminosity also rises at large invariant masses due to the transition matrix element updates, although well within the large uncertainties in this region. The quark-gluon luminosity shows a slight shape change upon addition of the FHMRUVV and TME updates, similar to that seen for the gluon-gluon luminosity, with an up to 2\% increase at $m_X \approx 400~{\rm GeV}$ reflecting the splitting function updates.

The impacts on phenomenology of approximate N3LO PDF updates will be analysed in more detail in the future as a complete update of the aN$^3$LO PDF determination is an ongoing study.

\section{Conclusion}\label{sec:conclusions}

In conclusion, we observe that upon updating the MSHT20 approximate N3LO PDFs with new information on the splitting functions and transition matrix elements there are only minor changes in the output PDFs, which are largely within uncertainties across the $x$ range. In turn the PDF luminosities show smaller modifications. The main changes concern a small increase in the gluon PDF in the Higgs region for the LHC, with a less than 2\% shift upwards from the updates. As a result, the gluon-gluon luminosity at $m_H$ changes by 2\%, around the upper edge of the original uncertainty band. This brings the results closer to those observed elsewhere. Overall this demonstrates the robustness of our original results and also serves as a posterior validation of our approximation procedure for the various theoretical ingredients and associated PDF MHOU uncertainties. We hope this therefore also serves as validation of the use of our approximate N3LO PDFs for LHC phenomenology.

\section*{Acknowledgements}
We would like to thank our colleagues in NNPDF, in particular Giacomo Magni, for providing the grids of updated transition matrix elements utilised in this work.

\paragraph{Funding information}
T.C. acknowledges
funding by Research Foundation-Flanders (FWO) (application number: 12E1323N and additional conference support via grant K121925N).
L. H.-
L. and R. S. T. thank STFC for support via grant awards ST/T000856/1 and ST/X000516/1.

\bibliographystyle{jhep}
\bibliography{references.bib}

\end{document}